\documentclass[ twocolumn,showpacs,superscriptaddress,aps,prl]{revtex4-1}  

\usepackage{graphicx}  
\usepackage{dcolumn}   
\usepackage{bm}        
\usepackage{amssymb}   

\usepackage{epsfig}
\usepackage[utf8]{inputenc}

\usepackage{color}
\usepackage[columnwise]{lineno}

\begin{document}

\title{
Experimental evidence of accelerated seismic release without critical failure in acoustic emissions of compressed nanoporous materials.
}
\author{Jordi Bar\'o}
\email{jordi.barourbea@ucalgary.ca}
\affiliation{Department of Physics, University of Illinois at Urbana Champaign,Urbana, Illinois 61801, USA.}
\affiliation{Department of Physics and Astronomy
University of Calgary.
2500 University Drive NW
Calgary, Alberta  T2N 1N4, Canada}
\affiliation{
Departament de F\'isica de la Mat\`eria Condensada. 
Facultat de F\'isica. Universitat de Barcelona. Mart\'i i Franqu\`es, 1. 
08028 Barcelona, Catalonia.
}
\author{Karin A. Dahmen}
\affiliation{Department of Physics, University of Illinois at Urbana Champaign,Urbana, Illinois 61801, USA.}

\author{Jörn Davidsen}

\affiliation{Department of Physics and Astronomy
University of Calgary.
2500 University Drive NW
Calgary, Alberta  T2N 1N4, Canada}

\author{Antoni Planes}
\affiliation{
Departament de F\'isica de la Mat\`eria Condensada. 
Facultat de F\'isica. Universitat de Barcelona. Mart\'i i Franqu\`es, 1. 
08028 Barcelona, Catalonia.
}
\author{Pedro O. Castillo}
\affiliation{
Departament de F\'isica de la Mat\`eria Condensada. 
Facultat de F\'isica. Universitat de Barcelona. Mart\'i i Franqu\`es, 1. 
08028 Barcelona, Catalonia.
}
\affiliation{
CONACYT, Instituto Tecnológico de Oaxaca, Av. Ing. Víctor Bravo Ahuja 125, Oaxaca de Juárez 68030, México
}

\author{Guillaume F. Nataf}
\affiliation{
Departament de F\'isica de la Mat\`eria Condensada. 
Facultat de F\'isica. Universitat de Barcelona. Mart\'i i Franqu\`es, 1. 
08028 Barcelona, Catalonia.
}
\affiliation{
Department of Materials Science, University of Cambridge, 27 Charles Babbage Road, Cambridge CB3 0FS, UK
}

\author{Ekhard K. H. Salje} 
\affiliation{Department of Earth Sciences, University of Cambridge,
Downing Street, Cambridge CB2 3EQ, UK.} 

\author{Eduard Vives}
\email{eduard@fmc.ub.edu}
\affiliation{
Departament de F\'isica de la Mat\`eria Condensada. 
Facultat de F\'isica. Universitat de Barcelona. Mart\'i i Franqu\`es, 1. 
08028 Barcelona, Catalonia.
}

\begin{abstract}
The total energy of acoustic emission (AE) events in externally stressed materials diverges when approaching macroscopic failure. Numerical and conceptual models explain this accelerated seismic release (ASR) as the approach to a critical point that coincides with ultimate failure. Here, we report ASR during soft uniaxial compression of three silica-based (SiO$_{2}$) nanoporous materials. Instead of a singular critical point, the distribution of AE energies is stationary and variations in the activity rate are sufficient to explain the presence of multiple periods of ASR leading to distinct brittle failure events. We propose that critical failure is suppressed in the AE statistics by mechanisms of transient hardening. Some of the critical exponents estimated from the experiments are compatible with mean field models, while others are still open to interpretation in terms of the solution of frictional and fracture avalanche models.
\end{abstract}


\maketitle

The mechanical deformation and failure of materials 
is a well documented case of avalanche dynamics \cite{Mogi1962,Davidsen2007,
Salje2011,Baro2013,Nataf2014,Nataf2014b,
Castillo2013,Salje2013,
Baro2016,Dahmen2009,Dahmen2011,
Hidalgo2002,Burridge1967,
Duan2015,
Sornette1992, Benzion1993, Dearcangelis1985,
Zapperi1997,Moreno2000,Amitrano2012,Alava2006,Shekhawat2013, Davidsen2017, Stanchits2006,Friedman2012,Maas2015,Antonaglia2014,Denisov2016, Rosti2010}. 
The energy of mechanical avalanches is partially released in elastic waves that can be detected by means of acoustic emission (AE) measurement \cite{Scruby1987}.
Several studies suggested the presence of a phase transition associated with the ultimate failure point \cite{Zapperi1997,Moreno2000,Amitrano2012,Alava2006, Benzion2008,Shekhawat2013} which could, in theory, be monitored and forecast by means of the statistical analysis of the preceding AE activity \cite{Sornette1995,Pradhan2002, Lippiello2012, Nataf2014b} and be used for hazard assessment. 
AE signals recorded during mechanical tests usually display a scale-free distribution of energies ($E$) close to a power-law: $D(E)dE \sim E^{-\varepsilon} dE$ with exponent $1 \lesssim \varepsilon \lesssim 2.5$.
Three different relationships are often reported between this scale-free phenomena and the proximity to failure:

({\it i}) The  exponent $\varepsilon$ in AE can decrease before failure~\cite{Scholz1968,Main1989,Amitrano2003,Goebel2013,Jiang2016,Jiang2017}.

({\it ii}) The rate of energy released over time diverges as a power-law with an exponent $m$ with respect to the time of failure $t_{c}$: 
\begin{equation}    
dE/dt (t)\propto (t_{c}-t)^{-m} ,
\label{eqASR}
\end{equation}
a phenomena called accelerated seismic release (ASR) in both seismology \cite{Jaume1999,Benzion2002} and also observed in AE experiments \cite{Main2000,Yin2004,Wang2008,Lennartz2014}.

 ({\it iii}) The characteristic scales of the avalanches depend on the distance to failure~\cite{Maas2015,Friedman2012,Antonaglia2014,Denisov2016}. This later observation supports the well established idea that failure occurs due to the divergence of correlation lengths at a critical point~\cite{Sornette1992,Mignan2011,Amitrano2012}. This so-called critical failure hypothesis predicts a generalized homogeneous distribution of event energies:
\begin{equation}
D(E;f)dE = E^{-\varepsilon} \mathcal{D}(E f ^{\beta}) dE 
= f^{\beta \varepsilon}\widetilde{\mathcal{D}}(Ef^{\beta})dE ,
\label{eqDE1}
\end{equation}
where ${\mathcal{D}}(x)$ and $\widetilde{\mathcal{D}}(x)$ are scaling functions, $f \equiv 1-t/t_{c}$ the time to failure and $\beta$ a characteristic exponent of the model.

While the exponent decrease ({\it i}) is currently not understood from a model perspective, 
 ASR ({\it ii}) and critical failure ({\it iii}) are well reproduced by most micromechanical models \cite{Sornette1992, Pradhan2002, Dearcangelis1985, Benzion1993,Mignan2011}. 
Since all statistical $n$-moments diverge at failure as $\langle E^{n} \rangle 
\sim f^{{(\varepsilon-1-n)}{\beta}}$ and the activity rate ($dN/dt$) is constant in most micromechanical models, ASR ({\it ii}) is a natural outcome of critical failure: 
\begin{equation}
dE/dt(f) = \langle E \rangle (f)  \: dN/dt(f) \sim f^{{(\varepsilon-2)}{\beta}}.
\label{eqASRcrit}
\end{equation}

Although ASR is assumed as a signature of criticality~\cite{Jaume1999,Mignan2011}, its connection with Eq.~(\ref{eqDE1}) is rarely tested with AE. 
Here, we analyze the AE during the approach to failure of nanoporous materials under soft uniaxial compression. We prove that ASR ({\it ii}) can appear in absence of progressive exponent changes ({\it i}) or critical failure ({\it iii}).
We estimate the experimental exponents $m$ (Eq.~(\ref{eqASR})), $\varepsilon$ (Eq.~(\ref{eqDE1})) and $\gamma$ relating the characteristic $E$ of an event with its duration $T$ through the conditional average: 
\begin{equation}
\langle E | T \rangle \propto T^{\gamma} ,
\label{eqProfile}
\end{equation} 
and interpret them in terms of the mean field solutions of fracture and frictional avalanches.

We limit our analysis to the three silica (SiO$_{2}$) based materials studied in Ref.~\cite{Nataf2014}:
natural red sandstone (SR2, $\Phi=17\%$ porosity) extracted from Arran Isle (UK) and two artificial porous silica glasses
Gelsil (Gel26, $\Phi=36\%$) and Vycor (V32, $\Phi=40\%$).  
Experimental details are found in Ref.~\cite{Nataf2014} and summarized in Table~\ref{tableExp}. 
Samples are compressed without lateral confinement at a steady quasistatically slow loading rate $dP/dt \sim  1$ kPa/s. 
The sample height ($h$) is measured over time with a laser extensometer and the AE is recorded by a piezoelectric transducer attached to the upper compression plate. Individual AE events are identified by thresholding the acoustic signal $V(t)$, defining the hitting time $t_{AE}$ and duration $D_{AE}$ of each AE event.
The AE energy of each event is computed as $E_{AE}\propto\int_{t_{AE}}^{t_{AE}+D_{AE}}\left|{V(t)}\right|^2dt$. 

\begin{table}[t]
\begin{tabular}{r  c c c c c}
    \hline
    &  area & height  & driving rate & $\mathit{Th}$ & $N$ \\
    &     $A$ (mm$^{2}$) &  $h$ (mm) &  $dP/dt$ (kPa/s) &(dB)& \\
    \hline
    \hline
Vycor (V32) & 17.0  & 5.65 & 5.7 & 23 & 34138 \\
Gelsil (G26)  & 46.7 & 6.2 & 0.7 & 26 & 5412 \\
Sands. (SR2)  & 17.0 & 4.3 & 2.4 & 23 & 27271\\
    \hline
\end{tabular} 
\caption{\label{tableExp} Sample details: crossectional area $A$; height $h$; compression rate $dP/dt$;  number $N$ of recorded signals above threshold {\it Th}.}   
\end{table}

\begin{figure}[]
\includegraphics[width=\columnwidth]{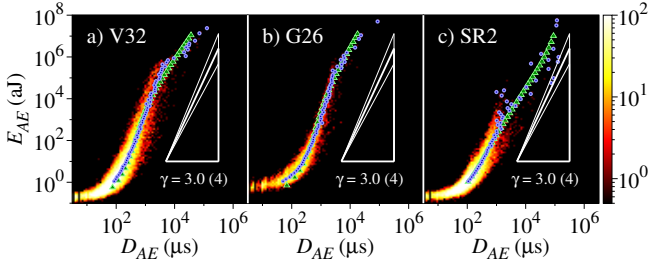}
\caption{\label{figMap}
(color online)
Histograms (color-coded) of AE events in the duration-energy ($D_{AE}$,$E_{AE}$) space. Blue dots: conditional averages  $\langle D_{AE}\rangle (E_{AE})$; green triangles: numerical solutions of $E_{AE} (D_{AE})$  consistent with Eq.~(\ref{eqProfile}) (see main text for details) with $\gamma=3.0(4)$ for V32, $\gamma=3.4(4)$ for G26 and $\gamma=3.2(4)$ for SR2.
}
\end{figure}

\begin{figure}[]
\includegraphics[width=1.0\columnwidth ]{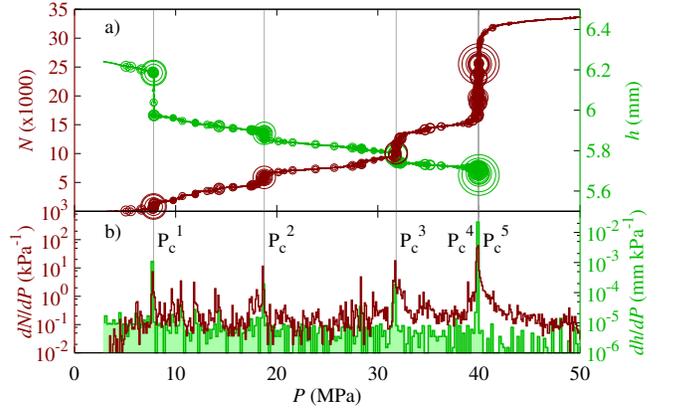}
\caption{\label{figSeq}
(color online)
 Mechanical response and AE sequence for experiment on Vycor (V32). 
(a) Cumulative number of events $N$ (dark-red) and height evolution $h$ (light-green) in experiment V32 as function of uniaxial pressure $P$. The size of the circles depends on the AE energy (size $ \sim E_{AE}^{0.25}$). 
(b) Mean AE activity rate $dN/dt$ (dark-red histograms) and strain rate $dh/dt$ (light-green histograms) in intervals of $\Delta P = 100$ kPa. Vertical gray lines: $P_{c}^{k}$.
}
\end{figure}

Fig.~\ref{figMap} shows the relations between AE energy ($E_{AE}$) and duration ($D_{AE}$)  in a density map, and the conditional averages $\langle D_{AE} \rangle (E_{AE})$. The experimental data is compared to a non-stochastic model considering a scale-free avalanche profile (Eq.~(\ref{eqProfile})) and the best value of $ \gamma $ found by inspection (see supplementary material).  Within error bars ($\pm 0.4$), all values are compatible with $\gamma=3$, as predicted by mean field (MF) models \cite{Dahmen2017,Baro2018}. The density clouds fill narrow stripes around the conditional average values as expected by Eq.~(\ref{eqProfile}). 

\begin{figure*}
\includegraphics[width=0.95\textwidth]{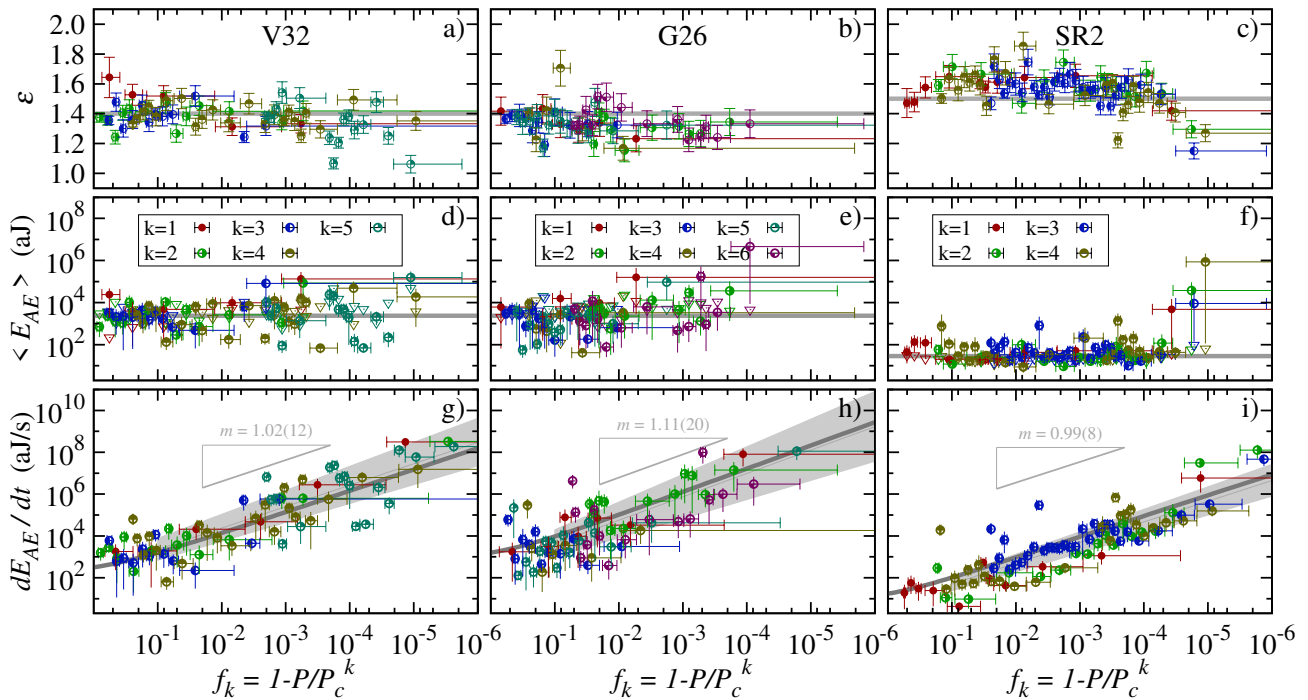}
\caption{\label{figScaling}(color online) 
Statistical variations with distance to strain drops $P_{c}^{k}$. The color-scheme identifies the index $k$.
 (a,b,c) Exponent $\widehat{\varepsilon}(f_{k})$ from Eq.~(\ref{eqDE1}) estimated within the interval ($1.0 - 1000$ aJ). 
 (d,e,f) Mean energy per signal $\langle E_{AE} \rangle(f_{k})$; expected mean value according to ${D}(E; E_{m},E_{c},\widehat{\varepsilon}(f_{k}))$ (triangles) with $E_{c}=10^{6}$ aJ ($10^{4}$ aJ for SR2); expected value from the global exponent (gray line).
 (g,h,i) Rate of AE energy $dE_{AE}/dt$;
 thin gray line: exponent $m$  fitted by least squares within $10^{-6}<f_{k}<10^{-1}$; thick gray line: a correction as expected by critical failure ( ${D}(E; E_{m},E_{m}f_{k}^{\beta(\widehat{\varepsilon},\widehat{m})},\widehat{\varepsilon}(f_{k}))$ with global $\widehat{\varepsilon}$ and estimated $\widehat{m}$. 
 The $f_{k}$ intervals of evaluation grow exponentially and have an imposed minimum size of $n=100$ signals ($n=50$ for G26).
 X-error bars: integration interval; Y-error bars: 90\% bootstrap interval in (d--i) and likelihood standard deviation in (d--f). Hard lower threshold imposed at $E_{m}=1.0$ aJ.}
\end{figure*}

The activity rate --- the number of AE events per time unit --- is non-stationary, as also reported in Refs.~\cite{Baro2013,Salje2013,Castillo2013,Nataf2014,Nataf2014b,Baro2016}. Fig.~\ref{figSeq}.a shows the mechanical evolution ($h(t)$) and the cumulative number of AE events ($N(t)$) for the experiment V32. Fig.~\ref{figSeq}.b shows the activity rate ($dN/dt$) and the decrease in height ($dh/dt$) evaluated in intervals of uniaxial pressure $\Delta P = 100$ kPa (converted from $t$ by $dP/dt$ in Table~\ref{tableExp}).
We identify several sharp drops in $h$ (five in Fig.~\ref{figSeq}), with a short characteristic temporal span $\Delta t_{c} \approx 0.1$~s (or  $\Delta P \approx 100$~Pa), at pressure values $P_{c}^{k}$. These so-called strain drops are outliers to an otherwise smooth strain evolution, as observed in the $dh/dP$ profile, and match a simultaneous increase of AE activity ($dN/dP$) and strong AE events. The events at $P_{c}^{k}$ resemble brittle failure, a typical outcome of internal weakening or progressive damage in MF micromechanical models \cite{Mehta2006,Dahmen2009}. 
Brittle failure events are macroscopic by definition. Thus, during a loading cycle a single (not multiple) brittle event is expected in these models. 
Here, however, the material recovers the stiffness during the intervals $P_{c}^{k}<P<P_{c}^{k+1}$ (Fig.~\ref{figSeq}). This can be explained by hardening, as reported in compression experiments \cite{Hidalgo2002}, due to the accommodation of the stress field. The presence of both weakening and hardening localizes damage in brittle events that can correspond to spallation correcting boundary defects \cite{Grady1988} or be arrested due to stress heterogeneities \cite{Laubie2017}. An ultimate failure event collapsing the whole sample is observed in all experiments ($P_{c}^{5}$ in Fig.~\ref{figSeq} has an associated $\Delta h \sim 5$ mm).

We study how the statistics of AE events are modified close to the most prominent stress drops by evaluating  $\langle E_{AE}\rangle$, $\varepsilon$ and $dE_{AE}/dt$ in short stress intervals correlated with the distance to each strain drop: $f_{k}:=1-P/P_{c}^{k}$. We select $P_{c}^{k}$ as the onset of each strain drop, identified with a precision of $0.01$s (equivalent to $\delta f_{k} \sim 10^{-6}-10^{-5}$) and compare the results to Eq.~(\ref{eqDE1}) where $\mathcal{D}$ is an exponential cutoff: 
\begin{equation}
{D}(E; E_{m},E_{c},\varepsilon) dE= 
E^{-\varepsilon}
\frac{
E_{c}^{\varepsilon-1}
\exp\left({-\frac{E}{E_{c}}}\right){}
}{
{\Gamma \left({1-\varepsilon,\frac{E_{m}}{E_{c}}}\right)}} dE.
\label{eqAnsatz}
 \end{equation}
Here, $\Gamma(a,x)$ is the incomplete gamma function and $E_{m}$ is the lower boundary of the distribution. $E_{c}$ is the characteristic scale of the exponential cutoff and, according to critical failure, should be proportional to $f_{k}^{-\beta}$ (Eq.~(\ref{eqDE1})). We truncate the distribution at the lower boundary $E_{m}=1$ aJ, to avoid  resolution artifacts distorting the power-law for low energies.

We inquire if the strain drops at $P_{c}^{k}$ can be interpreted as independent failure events, identified by at least one of the three trademarks mentioned earlier. 
Figs.~\ref{figScaling}.a--c show the exponents $\widehat{\varepsilon}(f_{k})$  estimated by Maximum Likelihood  inside the interval 1--1000 aJ \cite{Baro2012} (`$\:\widehat{~}\:$' denotes estimation), compared to the global estimated exponent (gray line). 
Figs.~\ref{figScaling}.d--f show the mean energy of individual AE events ($\langle E_{AE} \rangle (f_{k})$ in dots) compared to the solution to Eq.~(\ref{eqAnsatz}) (triangles) with  $\widehat{\varepsilon}(f_{k})$ from Figs.~\ref{figScaling}.a--c and stationary $\widehat{E}_{c}$ (gray lines). Lower panels (Figs.~\ref{figScaling}.g--i) show the rate of energy released by all events in temporal intervals ($dE_{AE}/dP(f_{k})$ in dots).  In Figs.~\ref{figScaling}.g--i, since some avalanches last longer than the evaluation intervals close to failure, their AE-energy is split in intervals of 1 ms in order to increase the temporal resolution.
The exponent $\widehat{\varepsilon}(f_{k})$ 
is almost stationary except for a few low values in the last intervals before $P_{c}^{k}$. Since all $\widehat{\varepsilon}(f_{k})<2$, critical failure expects a divergence in $\langle E_{AE} \rangle$ when $f_{k} \to 0$.
As first reported in Vycor~\cite{Baro2013}, $\langle E_{AE} \rangle(f_{k})$ is instead almost stationary and compatible with a finite and constant $\widehat{E}_{c}$ (see $E_{AE}$ distributions in supplementary material). 
Only the last intervals prior to failure show higher $\langle E_{AE} \rangle (f_{k})$, close to the 90\% confidence interval limit.  Despite the stationary $\langle E_{AE}\rangle$, all data sets exhibit a steady increase in $dE_{AE}/dt$ starting far from failure (Figs.~\ref{figScaling}.g--i), as predicted by ASR (Eq.~(\ref{eqASR})) considering $m \sim 1.0$ (thin gray lines). Thus, we observe ASR, even when avalanches are non-critical.

Fig.~\ref{figScaling} illustrates how ASR (Eq.~(\ref{eqASR})) is more general than critical failure (Eq.~(\ref{eqDE1})). This result can be reproduced by introducing microscopical mechanisms of transient hardening such as rheology damage~\cite{Lyakhovsky2005,Lyakhovsky2011}, rate-and-state dependent friction~\cite{Ruina1983} or viscoelasticity~\cite{Hainzl1999,Jagla2010,Lippiello2012}, into models that would otherwise exhibit critical failure~\cite{Mehta2006,Jagla2010,Baro2018}. 
Transient hardening acts as an effective dissipation \cite{Dahmen1998,Mehta2006,Baro2018} preventing criticality ~\cite{Vespignani1998,Lauritsen1996,Mehta2006,Jagla2010} and introduces temporal scales to the model reproducing the foreshock and aftershock sequences~\cite{Dieterich1979,Hainzl1999,Baro2018} also reported in our experiments~\cite{Baro2013,Nataf2014}.

Some of the last intervals preceding $P_{c}^{k}$ exhibit a significant decrease of $\widehat{\varepsilon}$ (see Fig.~\ref{figScaling}.c) and an increase in $\langle E_{AE} \rangle$ even higher than the expectation from Eq.~(\ref{eqAnsatz}) and the estimated $\widehat{\varepsilon}$ . 
Such intervals might contain superposition of events~\cite{Perez2004}, artifacts due to the signal clipping of large avalanches and/or strong AE related to brittle failure. As discussed in Ref.~\cite{Baro2018} brittle events can follow particular statistical laws.  Some experiments of rock fracture report instead a progressive decrease in $\widehat{\varepsilon}$ far from failure~\cite{Mogi1962, Scholz1968,Hirata1987,Lockner1993,Jiang2016} but this is not a universal feature~\cite{Lennartz2014} and it is also inconsistent with models~\cite{Amitrano2012}. 
Anisotropic stresses are known to affect  ${\varepsilon}$ in structural phase transitions~\cite{Niemann2012}, which might or might not play a role in rock fracture \cite{Lennartz2014}. The small size of our samples, close to the width of localization bands in sandstones~\cite{Baud2004,Lennartz2014}, might prevent any band-related anisotropy. 
Finally, several brittle events might commonly appear under uniaxial compression, since similar results were reported at constant stress~\cite{Salje2018}. Simulations can reproduce multifragmenation from dynamic fracture~\cite{Hild2003} or localized weakening bands in a predominantly hardening process~\cite{Amitrano2012, Duan2015}.

%
\begin{table}[t]
\begin{tabular}{| r l | c c c | c c |}
    \hline
    &   & V32  & G26 & SR2 & slip MF & fracture MF\\
    \hline
    \hline
    $\gamma$      &     & 3.0  (4) & 3.4  (4)  & 3.2  (4) & 3   &  3  \\
    $\varepsilon$ &     & 1.40 (5) & 1.40 (5)  & 1.50 (5) & 4/3 & 4/3 \\
    $m$           &     & 1.02 (13)& 1.11 (20) & 0.99 (8) & 1$^{{\it a}}$ 2$^{{\it b}}$   & 1/2$^{{\it a}}$ 1$^{{\it b}}$ \\
    \hline
    $\sigma \nu z$ & & 0.50 (6) & 0.45 (6) & 0.48 (5) & 1/2 & 1/2 \\ 
    $\kappa$       & & 1.60 (8) & 1.62 (8) & 1.76 (8) & 3/2 & 3/2 \\
    $\sigma^{{\it a}}$  & & 0.40 (9)  & 0.34 (9) & 0.24 (8) & 1/2 & 1 \\
    $\sigma^{{\it b}}$ & & 0.88 (12)  & 0.80 (16) & 0.76 (7) & 1/2 & 1 \\
    $\beta^{{\it a}}$   & & 3.7 $\pm$ 0.8   & 4.6 $\pm$ 1.2  & 6.3 $\pm 2.1$  & 3  & 3/2\\
    $\beta^{{\it b}}$  & & 1.67 (24)    & 1.83 (37)  & 2.00 (25)  & 3  & 3/2\\
    \hline
\end{tabular} 
\caption{\label{tableExpo} First three top rows: fitted exponents as represented in Fig.~\ref{figScaling}.g--i,  Fig.~\ref{figSeq}.a--c and Fig.~\ref{figScaling}.a--c, compared to the MF exponents for slip and fracture. Bottom rows: fundamental exponents estimated from MF theory. Superscripts $a$ (Eq.~(\ref{eqASRa})) and $b$ (Eq.~(\ref{eqASRcrit})) denote two different interpretations of ASR in terms of MF theory (see text).}   
\end{table}
%

Both friction and different fracture mechanisms are involved in  mechanical failure under compression~\cite{Stanchits2006,Fortin2006}. 
We compare the experimental values of $\varepsilon$,$\gamma$ and $m$ to the MF solutions of pure fracture and frictional models with transient hardening.  We consider the MF stick-slip model \cite{Dahmen2009,Dahmen2017} as a prototype for frictional avalanches and the democratic fiber bundle model \cite{Pradhan2002} for fracture (see supplementary material).
The collection of MF exponents \cite{Dahmen2009, Baro2018} is shown in Table~\ref{tableExpo}. 
 The critical exponents (Eqs.~\ref{eqDE1} and \ref{eqProfile}) are defined in terms of the size ($S$) of the avalanche from the relations:
\begin{equation}
    D(S;f)dS = S^{-\kappa} \mathcal{D}_{S}(S f ^{1/\sigma}) dS \quad;\quad \langle S | T \rangle \sim T^{{1}/{\sigma \nu z}} .
    \label{eqDS}
\end{equation}
In MF models the exponents $\kappa$, $\sigma \nu z$, $\varepsilon$ and  $\gamma$ are universal and invariant under transient hardening~\cite{Dahmen2009,Baro2018}. 
Given the broad regime with $\langle D_{AE} \rangle\sim E_{AE}^{{1}/{\gamma}}$ (Fig.~\ref{figMap}) we assume: $E_{AE} \propto E$. 
The estimated exponents $\varepsilon$ and $\gamma$ determine the values of $\kappa$ and $\sigma \nu z$, as shown in Table~\ref{tableExpo}. While $\sigma \nu z$ and $\beta$ are MF, $\kappa$ and $\varepsilon$ are higher but close to MF,  below 2-SD (standard deviation) in V32 and G26 and 3-SD in SR2, which might indicate the relevance of long-ranged elastic interactions.

The MF solutions of friction and fracture are similar, but differ in the values of $1/\sigma$ and $\beta$ related to the approach to failure  (see supplementary material).
Furthermore, the interpretation of $m$ in terms of the MF exponents is unclear when transient hardening is present.
According to MF models, the exponent $m$ defining the seismic energy released (Eq.~(\ref{eqASR})) is modified by transient hardening. 
Following Eq.~(\ref{eqDS}), the mean size in models with critical failure diverges as $\langle S \rangle (f)\sim f^\frac{\kappa-2}{\sigma}$ and, thus  $dS/dt\sim f^\frac{\kappa-2}{\sigma}$. Under slow driving, $dS/dt$ is invariant under transient hardening~\cite{Baro2018}. Considering the constant $\langle E \rangle (f)$ observed in Fig.~\ref{figScaling}.d--f, the MF model assumes that $\langle S \rangle (f)$ is also constant. Thus, $dS/dt$ diverges due to the divergence of $dN/dt$  and, instead of Eq.~(\ref{eqASRcrit}) we have:  
\begin{equation}
dE/dt(f) = \langle E \rangle(f)  \: dN/dt(f) \sim f^\frac{\kappa-2}{\sigma}.
\label{eqASRa}
\end{equation}
This interpretation of $dE/dt(f)$ derived from MF theory is presented with superscripts {\it a} in Table~\ref{tableExpo}. The experimental $m = (\kappa-2)/\sigma \approx 1$ coincides with the MF model of frictional avalanches. However, the values of $1/\sigma \sim 2.5-4$  and $\beta \sim 4-6$ are higher than the MF predictions of both models.

The relation between $m$ and  the fundamental exponents is discussed in MF theory, but not in models with local interactions, where transient hardening is known to affect the exponents ~\cite{Hainzl1999, Aragon2012}.  
An alternative hypothesis is that ASR (Eq.~(\ref{eqASRcrit})) is invariant under transient hardening. Then, $m = (\varepsilon-2)\beta \approx 1$ is compatible with the fracture MF model and the exponents $\sigma \sim 0.8$ and $\beta \sim 1.8$ are between both models, and notably closer to fracture (superscript {\it b} in Table~\ref{tableExpo}). The presence of brittle events denoting damage and related to fracture is consistent with this interpretation. Rock fracture experiments at low confining pressure \cite{Stanchits2006} are dominated by tensile fracture (not shear) AE events, a phenomena related to delitancy, and also reproduced in numerical simulations \cite{Camborde2000}.

In conclusion, sharp strain drops with massive AE events denoting brittle failure are identified during the compression of nanoporous materials. 
Instead of critical failure we find that $\langle E_{AE} \rangle$ is stationary and accelerated seismic release (ASR) is exclusively observed in the activity rate ($dN_{AE}/dt$). Previous experiments on sandstone under a different driving condition reported similar results ~\cite{Lennartz2014}.
Many theoretical models expect avalanche criticality at failure due to the divergence of correlation lengths~\cite{Sornette1992, Pradhan2002, Dearcangelis1985, Benzion1993,Mignan2011}. 
This criticality can be prevented by dissipation~\cite{Vespignani1998,Lauritsen1996,Jagla2010}, the dynamic weakening or hardening of the material ~\cite{Dahmen2009,Mehta2006} or the combined effect ~\cite{Dahmen1998}. 
In particular, the ASR and the lack of criticality reported here, together with the temporal correlations reported in Ref.~\cite{Nataf2014} can be reproduced by transient hardening~\cite{Baro2018}. In our experiment, an effective transient hardening can be caused by one or several internal micromechanical processes such as  
viscoelasticity \cite{Hainzl1999,Mainardi2011},  friction between crack surfaces \cite{Dieterich1979}, stress corrosion \cite{Bonamy2006}, diffusion of internal fluids \cite{Nur1972,Ehlers2000}, etc..
In contrast, externally measured slip avalanches usually scale to failure and appear unperturbed by transient hardening ~\cite{Friedman2012,Maas2015,Antonaglia2014,Denisov2016}. 
Analytic solutions of MF models allow us to interpret the experimental results in terms of critical exponents. While the interpretation of the ASR (Eq.~(\ref{eqASR}))  and its associated exponents remains an open question, other exponents are consistent with MF theory. 
 A remaining challenge for the future is to validate this extension of MF models to non-critical failure through new micromechanical experiments able to control the potential mechanisms of transient hardening and dissipation. \\

\begin{acknowledgments}
J.B. acknowledges the hospitality of the Department of Physics of the UIUC during a visit. We acknowledge fruitful discussion with M. LeBlanc. J.B., A.P and E.V. acknowledge financial support from the Spanish Ministry of Economy (MAT2016-75823-R). J.B and J.D. acknowledge financial support from NSERC.   KD gratefully acknowledges support from the National Science Foundation through grant NSF CBET-1336634, and thanks the KITP for hospitality and Support through grant NSF PHY-1125915.
\end{acknowledgments}


\begin{thebibliography}{10}

\bibitem{Mogi1962}
Kiyoo Mogi.
\newblock The influence of the dimensions of specimens on the fracture strength
  of rocks: comparison between the strength of rock specimens and that of the
  earth's crust.
\newblock 1962.

\bibitem{Davidsen2007}
J{\"o}rn Davidsen, Sergei Stanchits, and Georg Dresen.
\newblock Scaling and universality in rock fracture.
\newblock {\em Physical Review Letters}, 98(12):125502, 2007.

\bibitem{Salje2011}
Ekhard~KH Salje, Daniel~Enrique Soto-Parra, Antoni Planes, Eduard Vives, Marius
  Reinecker, and Wilfried Schranz.
\newblock Failure mechanism in porous materials under compression: crackling
  noise in mesoporous sio2.
\newblock {\em Philosophical Magazine Letters}, 91(8):554--560, 2011.

\bibitem{Baro2013}
Jordi Bar{\'o}, {\'A}lvaro Corral, Xavier Illa, Antoni Planes, Ekhard~KH Salje,
  Wilfried Schranz, Daniel~E Soto-Parra, and Eduard Vives.
\newblock Statistical similarity between the compression of a porous material
  and earthquakes.
\newblock {\em Physical Review Letters}, 110(8):088702, 2013.

\bibitem{Nataf2014}
Guillaume~F Nataf, Pedro~O Castillo-Villa, Jordi Bar{\'o}, Xavier Illa, Eduard
  Vives, Antoni Planes, and Ekhard~KH Salje.
\newblock Avalanches in compressed porous sio$_{2}$-based materials.
\newblock {\em Physical Review E}, 90(2):022405, 2014.

\bibitem{Nataf2014b}
Guillaume~F Nataf, Pedro~O Castillo-Villa, Pathikumar Sellappan, Waltraud~M
  Kriven, Eduard Vives, Antoni Planes, and Ekhard~KH Salje.
\newblock Predicting failure: Acoustic emission of berlinite under compression.
\newblock {\em Journal of Physics: Condensed Matter}, 26(27):275401, 2014.

\bibitem{Castillo2013}
Pedro~O Castillo-Villa, Jordi Bar{\'o}, Antoni Planes, Ekhard~KH Salje,
  Pathikumar Sellappan, Waltraud~M Kriven, and Eduard Vives.
\newblock Crackling noise during failure of alumina under compression: the
  effect of porosity.
\newblock {\em Journal of Physics: Condensed Matter}, 25(29):292202, 2013.

\bibitem{Salje2013}
Ekhard~KH Salje, Giulio~I Lampronti, Daniel~E Soto-Parra, Jordi Bar{\'o},
  Antoni Planes, and Eduard Vives.
\newblock Noise of collapsing minerals: Predictability of the compressional
  failure in goethite mines.
\newblock {\em American Mineralogist}, 98(4):609--615, 2013.

\bibitem{Baro2016}
Jordi Bar{\'o}, Antoni Planes, Ekhard~KH Salje, and Eduard Vives.
\newblock Fracking and labquakes.
\newblock {\em Philosophical Magazine}, 96(35):3686--3696, 2016.

\bibitem{Dahmen2009}
Karin~A Dahmen, Yehuda Ben-Zion, and Jonathan~T Uhl.
\newblock Micromechanical model for deformation in solids with universal
  predictions for stress-strain curves and slip avalanches.
\newblock {\em Physical Review Letters}, 102(17):175501, 2009.

\bibitem{Dahmen2011}
Karin~A Dahmen, Yehuda Ben-Zion, and Jonathan~T Uhl.
\newblock A simple analytic theory for the statistics of avalanches in sheared
  granular materials.
\newblock {\em Nature Physics}, 7(7):554, 2011.

\bibitem{Hidalgo2002}
Ra{\'u}l~Cruz Hidalgo, Christian~U Grosse, Ferenc Kun, Hans~W Reinhardt, and
  Hans~J Herrmann.
\newblock Evolution of percolating force chains in compressed granular media.
\newblock {\em Physical Review Letters}, 89(20):205501, 2002.

\bibitem{Burridge1967}
R~Burridge and Leon Knopoff.
\newblock Model and theoretical seismicity.
\newblock {\em Bulletin of the seismological society of america},
  57(3):341--371, 1967.

\bibitem{Duan2015}
K~Duan, CY~Kwok, and LG~Tham.
\newblock Micromechanical analysis of the failure process of brittle rock.
\newblock {\em International Journal for Numerical and Analytical Methods in
  Geomechanics}, 39(6):618--634, 2015.

\bibitem{Sornette1992}
Didier Sornette.
\newblock Mean-field solution of a block-spring model of earthquakes.
\newblock {\em Journal de Physique I}, 2(11):2089--2096, 1992.

\bibitem{Benzion1993}
Yehuda Ben-Zion and James~R Rice.
\newblock Earthquake failure sequences along a cellular fault zone in a
  three-dimensional elastic solid containing asperity and nonasperity regions.
\newblock {\em Journal of Geophysical Research: Solid Earth},
  98(B8):14109--14131, 1993.

\bibitem{Dearcangelis1985}
L~De~Arcangelis, S~Redner, and HJ~Herrmann.
\newblock A random fuse model for breaking processes.
\newblock {\em Journal de Physique Lettres}, 46(13):585--590, 1985.

\bibitem{Zapperi1997}
Stefano Zapperi, Purusattam Ray, H~Eugene Stanley, and Alessandro Vespignani.
\newblock First-order transition in the breakdown of disordered media.
\newblock {\em Physical Review Letters}, 78(8):1408, 1997.

\bibitem{Moreno2000}
Y~Moreno, JB~Gomez, and AF~Pacheco.
\newblock Fracture and second-order phase transitions.
\newblock {\em Physical Review Letters}, 85(14):2865, 2000.

\bibitem{Amitrano2012}
David Amitrano.
\newblock Variability in the power-law distributions of rupture events.
\newblock {\em The European Physical Journal-Special Topics}, 205(1):199--215,
  2012.

\bibitem{Alava2006}
Mikko~J Alava, Phani~KVV Nukala, and Stefano Zapperi.
\newblock Statistical models of fracture.
\newblock {\em Advances in Physics}, 55(3-4):349--476, 2006.

\bibitem{Shekhawat2013}
Ashivni Shekhawat, Stefano Zapperi, and James~P Sethna.
\newblock From damage percolation to crack nucleation through finite size
  criticality.
\newblock {\em Physical Review Letters}, 110(18):185505, 2013.

\bibitem{Davidsen2017}
J\"orn Davidsen, Grzegorz Kwiatek, Elli-Maria Charalampidou, Thomas Goebel,
  Sergei Stanchits, Marc R\"uck, and Georg Dresen.
\newblock Triggering processes in rock fracture.
\newblock {\em Physical Review Letters}, 119:068501, August 2017.

\bibitem{Stanchits2006}
Sergei Stanchits, Sergio Vinciguerra, and Georg Dresen.
\newblock Ultrasonic velocities, acoustic emission characteristics and crack
  damage of basalt and granite.
\newblock {\em Pure and Applied Geophysics}, 163(5-6):975--994, 2006.

\bibitem{Friedman2012}
Nir Friedman, Andrew~T Jennings, Georgios Tsekenis, Ju-Young Kim, Molei Tao,
  Jonathan~T Uhl, Julia~R Greer, and Karin~A Dahmen.
\newblock Statistics of dislocation slip avalanches in nanosized single
  crystals show tuned critical behavior predicted by a simple mean field model.
\newblock {\em Physical Review Letters}, 109(9):095507, 2012.

\bibitem{Maas2015}
R~Maa{\ss}, M~Wraith, JT~Uhl, JR~Greer, and KA~Dahmen.
\newblock Slip statistics of dislocation avalanches under different loading
  modes.
\newblock {\em Physical Review E}, 91(4):042403, 2015.

\bibitem{Antonaglia2014}
James Antonaglia, Xie Xie, Gregory Schwarz, Matthew Wraith, Junwei Qiao, Yong
  Zhang, Peter~K Liaw, Jonathan~T Uhl, and Karin~A Dahmen.
\newblock Tuned critical avalanche scaling in bulk metallic glasses.
\newblock {\em Scientific Reports}, 4, 2014.

\bibitem{Denisov2016}
DV~Denisov, KA~L{\"o}rincz, JT~Uhl, KA~Dahmen, and P~Schall.
\newblock Universality of slip avalanches in flowing granular matter.
\newblock {\em Nature communications}, 7:10641, 2016.

\bibitem{Rosti2010}
J~Rosti, J~Koivisto, and MJ~Alava.
\newblock Statistics of acoustic emission in paper fracture: precursors and
  criticality.
\newblock {\em Journal of Statistical Mechanics: Theory and Experiment},
  2010(02):P02016, 2010.

\bibitem{Scruby1987}
CB~Scruby.
\newblock An introduction to acoustic emission.
\newblock {\em Journal of Physics E: Scientific Instruments}, 20(8):946, 1987.

\bibitem{Benzion2008}
Yehuda Ben-Zion.
\newblock Collective behavior of earthquakes and faults: Continuum-discrete
  transitions, progressive evolutionary changes, and different dynamic regimes.
\newblock {\em Reviews of Geophysics}, 46(4), 2008.

\bibitem{Sornette1995}
Didier Sornette and Charles~G Sammis.
\newblock Complex critical exponents from renormalization group theory of
  earthquakes: Implications for earthquake predictions.
\newblock {\em Journal de Physique I}, 5(5):607--619, 1995.

\bibitem{Pradhan2002}
Srutarshi Pradhan and Bikas~K Chakrabarti.
\newblock Precursors of catastrophe in the bak-tang-wiesenfeld, manna, and
  random-fiber-bundle models of failure.
\newblock {\em Physical Review E}, 65(1):016113, 2001.

\bibitem{Lippiello2012}
Eugenio Lippiello, Warner Marzocchi, L~De~Arcangelis, and C~Godano.
\newblock Spatial organization of foreshocks as a tool to forecast large
  earthquakes.
\newblock {\em Scientific Reports}, 2, 2012.

\bibitem{Scholz1968}
CH~Scholz.
\newblock The frequency-magnitude relation of microfracturing in rock and its
  relation to earthquakes.
\newblock {\em Bulletin of the Seismological Society of America},
  58(1):399--415, 1968.

\bibitem{Main1989}
Ian~G Main, Philip~G Meredith, and Colin Jones.
\newblock A reinterpretation of the precursory seismic b-value anomaly from
  fracture mechanics.
\newblock {\em Geophysical Journal International}, 96(1):131--138, 1989.

\bibitem{Amitrano2003}
David Amitrano.
\newblock Brittle-ductile transition and associated seismicity: Experimental
  and numerical studies and relationship with the b value.
\newblock {\em Journal of Geophysical Research: Solid Earth}, 108(B1), 2003.

\bibitem{Goebel2013}
TH~W~Goebel, D~Schorlemmer, TW~Becker, G~Dresen, and CG~Sammis.
\newblock Acoustic emissions document stress changes over many seismic cycles
  in stick-slip experiments.
\newblock {\em Geophysical Research Letters}, 40(10):2049--2054, 2013.

\bibitem{Jiang2016}
Xiang Jiang, Deyi Jiang, Jie Chen, and Ekhard~KH Salje.
\newblock Collapsing minerals: Crackling noise of sandstone and coal, and the
  predictability of mining accidents.
\newblock {\em American Mineralogist}, 101(12):2751--2758, 2016.

\bibitem{Jiang2017}
Xiang Jiang, Hanlong Liu, Ian~G Main, and Ekhard~KH Salje.
\newblock Predicting mining collapse: Superjerks and the appearance of
  record-breaking events in coal as collapse precursors.
\newblock {\em Physical Review E}, 96(2):023004, 2017.

\bibitem{Jaume1999}
Steven~C Jaum{\'e} and Lynn~R Sykes.
\newblock Evolving towards a critical point: A review of accelerating seismic
  moment/energy release prior to large and great earthquakes.
\newblock In {\em Seismicity Patterns, their Statistical Significance and
  Physical Meaning}, pages 279--305. Springer, 1999.

\bibitem{Benzion2002}
Yehuda Ben-Zion and Vladimir Lyakhovsky.
\newblock Accelerated seismic release and related aspects of seismicity
  patterns on earthquake faults.
\newblock {\em Earthquake Processes: Physical Modelling, Numerical Simulation
  and Data Analysis Part II}, pages 2385--2412, 2002.

\bibitem{Main2000}
Ian~G Main.
\newblock A damage mechanics model for power-law creep and earthquake
  aftershock and foreshock sequences.
\newblock {\em Geophysical Journal International}, 142(1):151--161, 2000.

\bibitem{Yin2004}
Xiang-chu Yin, Huai-zhong Yu, Victor Kukshenko, Zhao-Yong Xu, Zhishen Wu, Min
  Li, Keyin Peng, Surgey Elizarov, and Qi~Li.
\newblock Load-unload response ratio (lurr), accelerating moment/energy release
  ({AM}/er) and state vector saltation as precursors to failure of rock
  specimens.
\newblock In {\em Computational Earthquake Science Part II}, pages 2405--2416.
  Springer, 2004.

\bibitem{Wang2008}
Lifeng Wang, Shengli Ma, and Li~Ma.
\newblock Accelerating moment release of acoustic emission during rock
  deformation in the laboratory.
\newblock {\em Pure and Applied Geophysics}, 165(2):181--199, 2008.

\bibitem{Lennartz2014}
S~Lennartz-Sassinek, IG~Main, M~Zaiser, and CC~Graham.
\newblock Acceleration and localization of subcritical crack growth in a
  natural composite material.
\newblock {\em Physical Review E}, 90(5):052401, 2014.

\bibitem{Mignan2011}
Arnaud Mignan.
\newblock Retrospective on the accelerating seismic release (asr) hypothesis:
  Controversy and new horizons.
\newblock {\em Tectonophysics}, 505(1):1--16, 2011.

\bibitem{Dahmen2017}
Karin~A Dahmen.
\newblock Mean field theory of slip statistics.
\newblock In {\em Avalanches in Functional Materials and Geophysics}, pages
  19--30. Springer, 2017.

\bibitem{Baro2018}
Jordi Bar\'o and J\"orn Davidsen.
\newblock Universal avalanche statistics and triggering close to failure in a
  mean field model of rheological fracture.
\newblock {\em arXiv preprint arXiv:1801.01930 (accepted for publication in PRE)}, 2018.

\bibitem{Mehta2006}
Amit~P Mehta, Karin~A Dahmen, and Yehuda Ben-Zion.
\newblock Universal mean moment rate profiles of earthquake ruptures.
\newblock {\em Physical Review E}, 73(5):056104, 2006.

\bibitem{Grady1988}
DE~Grady.
\newblock The spall strength of condensed matter.
\newblock {\em Journal of the Mechanics and Physics of Solids}, 36(3):353--384,
  1988.

\bibitem{Laubie2017}
Hadrien Laubie, Farhang Radjai, Roland Pellenq, and Franz-Josef Ulm.
\newblock Stress transmission and failure in disordered porous media.
\newblock {\em Physical Review Letters}, 119(7):075501, 2017.

\bibitem{Baro2012}
Jordi Bar{\'o} and Eduard Vives.
\newblock Analysis of power-law exponents by maximum-likelihood maps.
\newblock {\em Physical Review E}, 85(6):066121, 2012.

\bibitem{Lyakhovsky2005}
Vladimir Lyakhovsky, Yehuda Ben-Zion, and Amotz Agnon.
\newblock A viscoelastic damage rheology and rate-and state-dependent friction.
\newblock {\em Geophysical Journal International}, 161(1):179--190, 2005.

\bibitem{Lyakhovsky2011}
Vladimir Lyakhovsky, Yariv Hamiel, and Yehuda Ben-Zion.
\newblock A non-local visco-elastic damage model and dynamic fracturing.
\newblock {\em Journal of the Mechanics and Physics of Solids},
  59(9):1752--1776, 2011.

\bibitem{Ruina1983}
Andy Ruina.
\newblock Slip instability and state variable friction laws.
\newblock {\em Journal of Geophysical Research: Solid Earth},
  88(B12):10359--10370, 1983.

\bibitem{Hainzl1999}
Sebastian Hainzl, Gert Z{\"o}ller, and J{\"u}rgen Kurths.
\newblock Similar power laws for foreshock and aftershock sequences in a
  spring-block model for earthquakes.
\newblock {\em Journal of Geophysical Research: Solid Earth},
  104(B4):7243--7253, 1999.

\bibitem{Jagla2010}
EA~Jagla and AB~Kolton.
\newblock A mechanism for spatial and temporal earthquake clustering.
\newblock {\em Journal of Geophysical Research: Solid Earth}, 115(B5), 2010.

\bibitem{Dahmen1998}
Karin Dahmen, Deniz Erta{\c{s}}, and Yehuda Ben-Zion.
\newblock Gutenberg-richter and characteristic earthquake behavior in simple
  mean-field models of heterogeneous faults.
\newblock {\em Physical Review E}, 58(2):1494, 1998.

\bibitem{Vespignani1998}
Alessandro Vespignani and Stefano Zapperi.
\newblock How self-organized criticality works: A unified mean-field picture.
\newblock {\em Physical Review E}, 57(6):6345, 1998.

\bibitem{Lauritsen1996}
Kent~B{\ae}kgaard Lauritsen, Stefano Zapperi, and H~Eugene Stanley.
\newblock Self-organized branching processes: Avalanche models with
  dissipation.
\newblock {\em Physical Review E}, 54(3):2483, 1996.

\bibitem{Dieterich1979}
James~H Dieterich.
\newblock Modeling of rock friction: 1. experimental results and constitutive
  equations.
\newblock {\em Journal of Geophysical Research: Solid Earth},
  84(B5):2161--2168, 1979.

\bibitem{Perez2004}
Francisco-Jos{\'e} P{\'e}rez-Reche, Bosiljka Tadi{\'c}, Llu{\'\i}s Ma{\~n}osa,
  Antoni Planes, and Eduard Vives.
\newblock Driving rate effects in avalanche-mediated first-order phase
  transitions.
\newblock {\em Physical Review Letters}, 93(19):195701, 2004.

\bibitem{Hirata1987}
Takayuki Hirata.
\newblock Omori's power law aftershock sequences of microfracturing in rock
  fracture experiment.
\newblock {\em Journal of Geophysical Research: Solid Earth},
  92(B7):6215--6221, 1987.

\bibitem{Lockner1993}
D~Lockner.
\newblock The role of acoustic emission in the study of rock fracture.
\newblock In {\em International Journal of Rock Mechanics and Mining Sciences
  \& Geomechanics Abstracts}, volume~30, pages 883--899. Elsevier, 1993.

\bibitem{Niemann2012}
R~Niemann, J~Bar{\'o}, O~Heczko, L~Schultz, S~F{\"a}hler, E~Vives,
  Ll~Ma{\~n}osa, and A~Planes.
\newblock Tuning avalanche criticality: Acoustic emission during the
  martensitic transformation of a compressed ni-mn-ga single crystal.
\newblock {\em Physical Review B}, 86(21):214101, 2012.

\bibitem{Baud2004}
Patrick Baud, Emmanuelle Klein, and Teng-fong Wong.
\newblock Compaction localization in porous sandstones: spatial evolution of
  damage and acoustic emission activity.
\newblock {\em Journal of Structural Geology}, 26(4):603--624, 2004.

\bibitem{Salje2018}
Ekhard~KH Salje, Hanlong Liu, Linsen Jin, Deyi Jiang, Yang Xiao, and Xiang
  Jiang.
\newblock Intermittent flow under constant forcing: Acoustic emission from
  creep avalanches.
\newblock {\em Applied Physics Letters}, 112(5):054101, 2018.

\bibitem{Hild2003}
Fran{\c{c}}ois Hild, Christophe Denoual, Pascal Forquin, and Xavier Brajer.
\newblock On the probabilistic--deterministic transition involved in a
  fragmentation process of brittle materials.
\newblock {\em Computers \& Structures}, 81(12):1241--1253, 2003.

\bibitem{Fortin2006}
J{\'e}r{\^o}me Fortin, Sergei Stanchits, Georg Dresen, and Yves Gu{\'e}guen.
\newblock Acoustic emission and velocities associated with the formation of
  compaction bands in sandstone.
\newblock {\em Journal of Geophysical Research: Solid Earth}, 111(B10), 2006.

\bibitem{Aragon2012}
LE~Arag{\'o}n, EA~Jagla, and A~Rosso.
\newblock Seismic cycles, size of the largest events, and the avalanche size
  distribution in a model of seismicity.
\newblock {\em Physical Review E}, 85(4):046112, 2012.

\bibitem{Camborde2000}
F~Camborde, C~Mariotti, and FV~Donz{\'e}.
\newblock Numerical study of rock and concrete behaviour by discrete element
  modelling.
\newblock {\em Computers and Geotechnics}, 27(4):225--247, 2000.

\bibitem{Mainardi2011}
Francesco Mainardi and Giorgio Spada.
\newblock Creep, relaxation and viscosity properties for basic fractional
  models in rheology.
\newblock {\em The European Physical Journal-Special Topics}, 193(1):133--160,
  2011.

\bibitem{Bonamy2006}
Daniel Bonamy, Silke Prades, CL~Rountree, Laurent Ponson, Davy Dalmas,
  Elisabeth Bouchaud, K~Ravi-Chandar, and Claude Guillot.
\newblock Nanoscale damage during fracture in silica glass.
\newblock {\em International Journal of Fracture}, 140(1):3--14, 2006.

\bibitem{Nur1972}
Amos Nur and John~R Booker.
\newblock Aftershocks caused by pore fluid flow?
\newblock {\em Science}, 175(4024):885--887, 1972.

\bibitem{Ehlers2000}
W~Ehlers and B~Markert.
\newblock On the viscoelastic behaviour of fluid-saturated porous materials.
\newblock {\em Granular Matter}, 2(3):153--161, 2000.

\end{thebibliography}



\end{document}